\documentclass[12pt]{article}

\usepackage{graphicx}                      
\usepackage{epstopdf}
\usepackage{amsmath} 
\usepackage{braket}
\usepackage{xcolor}
\usepackage{hyperref}
\usepackage{csquotes}

\usepackage{amsmath,amssymb,amsfonts,color,graphicx,color,soul}
\usepackage{epsfig}
\usepackage{blindtext}
\usepackage[utf8]{inputenc}
\usepackage{caption}
\usepackage{subcaption}
\usepackage{capt-of}
\usepackage{tabu}
\usepackage{hyperref}
\usepackage{csquotes}
\usepackage{amsmath,amsfonts,amssymb}
\usepackage{graphicx}
\usepackage{latexsym,amstext,amscd}
\usepackage{setspace}
\usepackage{epic}
\usepackage{mathrsfs}
\usepackage[T1]{fontenc}
\usepackage[T1]{fontenc}
\usepackage[utf8]{inputenc}

\usepackage[backend=biber,
  sorting=none, 
  url=true,                  
  hyperref=true,
  style=numeric-comp]{biblatex}
\evensidemargin \oddsidemargin
\allowdisplaybreaks

\usepackage[bottom]{footmisc}
\usepackage[affil-it]{authblk}
\usepackage[english]{babel}
\usepackage{dcolumn}

\title{Taming the long distance effects in the $D^{+}_{s}\rightarrow \pi^{+}\ell^{-}\ell^{+}$ decay}
\author{Marxil S\'anchez }
\author{Genaro Toledo}
\affil{Instituto de F\1sica, Universidad Nacional Aut\'onoma de M\'exico, AP 20-364, Ciudad de M\'exico 01000, M\'exico.}
\author{I. Heredia~De~La~Cruz}
\affil{Consejo Nacional de Ciencia y Tecnolog\'ia, M\'exico}
\affil{Departamento de F\1sica, Centro de Investigaci\'on y de Estudios Avanzados del IPN, M\'exico.}

\date{\today}

\begin{document}

\maketitle
\begin{abstract}
 
The semileptonic decay $D^{+}_{s}\rightarrow \pi^{+}\ell^{-}\ell^{+}$, where  $\ell=e, \ \mu$, is used as a reference channel when looking for suppressed or forbidden processes of the standard model of the form $D_{s}\rightarrow P\ell^{\pm}\ell^{\prime+}$. The process is dominated by hadronic resonances from the $\eta$ to the $\phi$ meson region, usually excluded due to the large uncertainties associated. In this work,  we explore the role of the parameters involved in the resonance region. We focus in the relative strong phase between the $\rho$ and the $\phi$ meson ($\delta_{\rho\phi}$). We argue that this parameter can be bound from LHCb data of the di-muon invariant mass in the $D^{+}_{s}\rightarrow \pi^{+}\mu^{-}\mu^{+}$ decay and we perform a fit to provide a first approach to it. We obtain $\delta_{\rho\phi}=(0.44 \pm 0.24) \pi$. In contradistinction, it would be useful to consider observables insensitive to this phase. We first compute the invariant mass of the lepton pair at a given angle of one of the leptons emission with respect to the pion, and then we compute the forward-backward distribution at such angle. We show that for the latter observable the resonance region is smeared, exhibiting no dependence on the $\delta_{\rho\phi}$ phase, although global effects are observed by comparing with the pure phase space approach. In order to consider this observable in more general scenarios, we analyse the behavior for the resonant background process $D^{+}_{s}\rightarrow \pi\pi\pi$, the short distance contribution in the standard model, and the charged Higgs mediated process as given by the two Higgs doublet model, which exhibit distinguishable features among themselves.
\end{abstract}

\section{Introduction}
The advent of more precise measurements in the charm sector will allow to test both hadronic models and new physics scenarios in an intermediate energy region~\cite{Gisbert:2020vjx,deBoer:2015boa,Bause:2019vpr,Fajfer:2015mia,Burdman:2001tf,Guo:2017edd,Wang:2014uiz,Fajfer:2012nr,Tahir:2011npe,Paul:2011ar}. In the hadronic side, the effective low-energy approaches and the heavy quark approximation are taken to the corner of their validity region. The new physics scenarios that may be at reach in the charm sector require a proper account of the different hadronic decay modes, whose contribution are dominant and usually taken as reference when looking for suppressed or forbidden decays of the standard model (SM).
In particular, the $D^{+}_{s}\rightarrow \pi^{+}\ell^{-}\ell^{+}$ decay has been used as a reference channel in the understanding of suppressed modes involving flavor changing neutral currents (FCNC), induced at the one loop level~\cite{LHCb:2013hxr}. Although this decay does not proceed through that mode, it serves as a baseline upon the inclusion of the FCNC channels (as in the $D_{s}\rightarrow K^{+}\ell^{+}\ell^{-}$ decay). Namely, knowing better the full aspects of non-FCNC processes would help to identify the truly FCNC contribution features.  Moreover, the new physics prospects in the charm sector are expected to be tested in the so-called off-resonance region, taken roughly below the $\eta$ and above the $\phi$ masses~\cite{Gisbert:2020vjx,deBoer:2015boa,Bause:2019vpr,Fajfer:2015mia,LHCb:2013hxr}. The intermediate region is cumbersome, despite including well known resonances, mainly because the relative strong phases are not settled.   Reported measurements \cite{LHCb:2013hxr} do not consider the full kinematical region to set the constrains on the processes involving FCNC contributions; rather, they exclude the resonant region and replace it by a phase space based approach. That is, the constrains do not rest upon a proper understanding of this region, but on the phase space approach. While this is a way to avoid this by far cumbersome region, a better knowledge of all the physical aspects involved would allow to set constraints with fewer assumptions.\\

Thus, in this work, we revisit the long distance (LD) description of the $D^{+}_{s}\rightarrow \pi^{+}\ell^{-}\ell^{+}$ decay. The hadronic parameters involved are computed in the meson dominance approach \cite{Bando:1984ej,Meissner:1987ge} and compared with other estimates. We focus in the relative phase between the $\rho$ and the $\phi$ mesons, and point out that this phase can be bound from the di-muon invariant mass distribution in the $D^{+}_{s}\rightarrow \pi^{+}\mu^{-}\mu^{+}$ decay, which has been measured by the LHCb experiment \cite{LHCb:2013hxr}. The experiment does not provide the details on the resonances analysis, which should include the corresponding phase. Therefore, in order to illustrate that the available data are already sensitive to this parameter, we have computed the di-lepton invariant mass and performed a $\chi^2$ fit in the region between the $\rho$ and the $\phi$ resonances. The obtained value should be considered as an indication of the experimental feasibility to determine this parameter accurately with a proper handling of the data, incorporating the detector effects.
On the other hand, earlier works have explored angular observables~\cite{deBoer:2015boa,Gisbert:2020vjx,Bause:2019vpr,Fajfer:2015mia,Sahoo:2017lzi,Bharucha:2020eup,Burdman:2001tf,Paul:2011ar} to distinguish different properties of the system under study. Analyses of hadronic data using chiral approaches and general properties like unitarity and analiticity have also improved the understanding of the system~\cite{Bharucha:2020eup}. Observables insensitive to the relative phase would be useful to keep the hadronic contributions under control. For that purpose, here, we compute the di-lepton invariant mass, for a given angle of one of the leptons emission with respect to the pion, and then we compute the corresponding forward-backward distribution. The latter observable is found to exhibit no dependency on the relative phase in the full kinematical region; a comparison with  the pure phase space approach is used as a baseline. 
In order to consider this observable in more general scenarios, that would allow the experimentalists to distinguish not only the LD but also the short distance (SD) and non-standard contributions, we analyse the behavior for the resonant background process $D^{+}_{s}\rightarrow \pi\pi\pi$~\cite{FOCUS:2003tdy}, SD contribution in the SM, and the charged Higgs mediated process as given by the Type-II two-Higgs-doublet model (2HDM-II)~\cite{Hou:1992sy,Branco:2011iw}, to explore their features. Finally, we discuss our results.
  
\begin{figure}
\includegraphics[width=15cm]{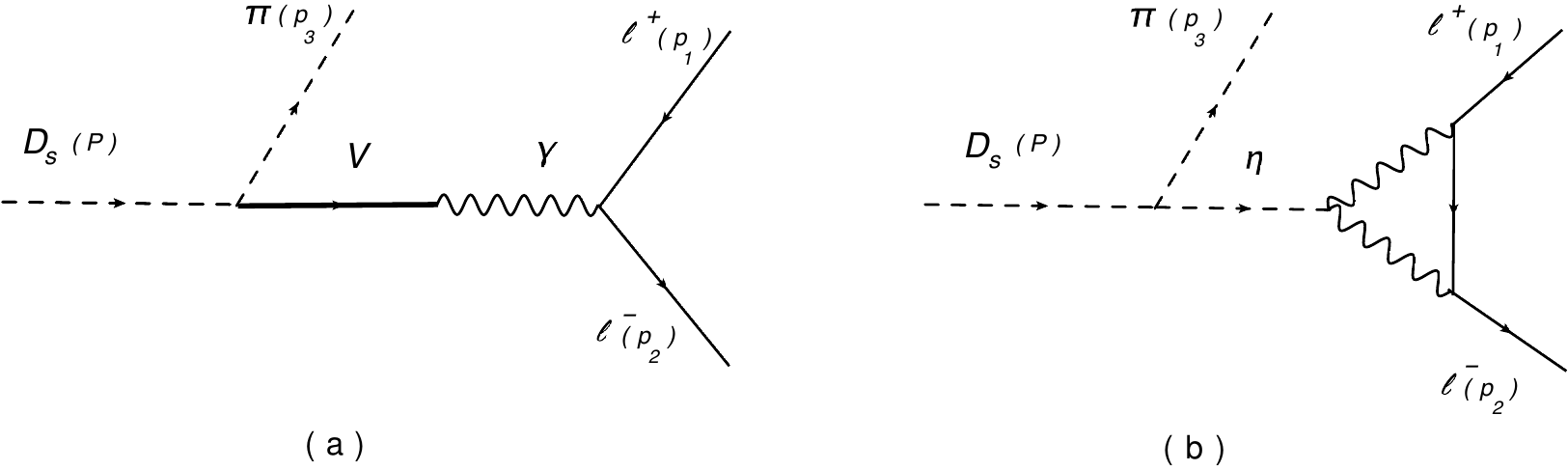}
\caption{Long distance (LD) contributions to the $D_s \to \pi \ell^{-}\ell^{+}$ decay, driven by vector (a) and  pseudo-scalar (b) mesons.}
\label{Fig:1}
\end{figure}

\section{The $D^{+}_{s}\to \pi^{+}\ell^{-}\ell^{+}$ decay}
The LD contribution to the $D^{+}_{s}\rightarrow \pi^{+}\ell^{-}\ell^{+}$ decay is induced by an effective non-leptonic weak Lagrangian 
\begin{equation}
    \mathcal{L}^{|\Delta c|=1} \propto -\frac{G_F}{\sqrt{2}}V^*_{c{q_j}}V_{uq_i}[\bar{u}\gamma^\mu(1-\gamma^5)q_i\bar{q}_j\gamma^\mu(1-\gamma^5)c+ ...],
    \label{weakl}
\end{equation}
where $q_{i,j}$ denote $d$ or $s$ quarks, and the proportionality factor can be determined from experimental information \cite{Fajfer:2001sa}. The factorization hypothesis is invoked to determine the corresponding matrix elements. We follow this approximation which, as we will show later for this particular case, is expected to hold at the current experimental precision although it may become relevant in a more detailed analysis. Deviations from this approximation are important in processes involving FCNC \cite{Feldmann:2017izn,Bharucha:2020eup,Beylich:2011aq,Grinstein:2004vb}.
Therefore, the LD part of the $D^{+}_{s}\rightarrow \pi^{+}\ell^{-}\ell^{+}$ decay can be seen as the effective transition from $D_s$ to $\pi$, driven by vector and pseudo-scalar structures, and then the production of the lepton pair~\cite{deBoer:2015boa,Burdman:2001tf,Fajfer:2001sa}, as depicted in Fig.~\ref{Fig:1}. The process is dominated by intermediate vector mesons ($V$) leading to the production of the lepton pair. In this case, the amplitude in the  meson dominance approach, using the momenta assignment $D^{+}_{s}(P)\rightarrow \ell^{+}(p_1) \ell^{-}(p_2)  \pi^{+}(p_3)$, is given by
\begin{equation}
\mathcal{M}_V= i \frac{e^2  G_{D_s\pi V}} {g_{V}}\frac{(P+p_3)_\nu l^\nu}{k^2-m_V^2+im_V\Gamma_V}\;, 
\label{mvector}
\end{equation}
where $k^2=(p_1+p_2)^2$ is the square of the di-lepton invariant mass, $l^\nu \equiv \bar{u}(p_2)\gamma^\nu \nu(p_1)$ is the leptonic current, $m_V$ and $\Gamma_V$ are the mass and total width of the vector meson, $g_{V}$ is the vector-photon coupling, and $G_{D_s\pi V}$ is the effective coupling of the $D_s -\pi$ to the vector meson (the ordering of the sub-indices indicates the decay mode), which includes a factor $G_F V^*_{cs}V_{ud}/\sqrt{2}$ associated to the weak transitions and the energy dependence factor $m_V^2/k^2$.
The $G_{D_s\pi V}$ and $g_{V}$ couplings can be estimated from the measurement of the branching ratio of the two-body decays $D_{s}\rightarrow \pi V$ and $V\rightarrow \ell^{+}\ell^{-}$, respectively \cite{ParticleDataGroup:2020ssz}.

In general, a $G_{P_{1}P_{2}V}$ coupling can be extracted from a $P_1\to P_{2}V$ decay, with $P_{1,2}$ pseudo-scalar mesons, whose effective amplitude can be written as
\begin{equation}
\mathcal{M}=i G_{P_{1}P_{2}V}(p_{1}+p_{2})^{\mu}\eta^{\ast}_{\mu}(q),
\end{equation}
where $p_{1}$, $p_{2}$ and $q$ are the corresponding momenta and $\eta^{\ast}_{\mu}$ is the vector meson polarization tensor. The coupling is given by
\begin{eqnarray}
G_{P_{1}P_{2}V}=\Big (\frac{16\pi m^{2}_{V}m^{3}_{P_{1}}\Gamma_{P_{1}P_{2}V}}{\lambda^{3/2}(m_{P_{1}}^{2},m_{P_{2}}^{2},m_{V}^{2})} \Big )^{1/2},
\end{eqnarray}
where $\lambda (x,y,z) =x^2+y^2+z^2-2xy-2xz-2yz$ and $\Gamma_{P_1P_2V}$ is the corresponding decay width~\cite{ParticleDataGroup:2020ssz}. $G_{D_{s}\pi V}$ can be also obtained from three-body decays of the form $D_s \to \pi V \to \pi P_1P_2$ which, in addition, involves the $G_{VP_{1}P_{2}}$ coupling. A deviation between the two ways to extract $G_{D_{s}\pi V}$ could give an indication of the effect of the energy dependence and/or factorization approach. We perform such comparison whenever there are available data and find no deviation from each other within the current experimental uncertainties.

In Table~\ref{couplings1}, we show the values for the $G_{D_{s}\pi V}$ couplings ($V=\rho, \ \omega, \ \phi$). For completeness, we also include the values for other vector mesons whether or not they contribute to our leptonic decay mode, provided the experimental information is available. The coupling involving the $\rho(1450)$ meson is estimated considering the full $D_s \to \pi V \to 3\pi$ decay. For comparison, values are also quoted for the $D_s \to K V$ decay, although the internal decay mechanism is different~\cite{deBoer:2015boa,Fajfer:2015mia,Burdman:2001tf}.

\begin{table}[t]
\begin{center}
\begin{tabular}{|c||c|c|c|c|c|c|}
\hline
 $V  \to $&  $\rho$& $\omega$& $\phi$& $\rho(1450)$& $K^{\ast} (892)$& $K^{\ast} (1410)$\\
\hline
 $D_{s}^{+}\rightarrow \pi^{+} V$& $4.13\times 10^{-8}$&  $1.32\times 10^{-7}$& $1.03\times 10^{-6}$
 & $7.91\times 10^{-7}$&  $1.40\times 10^{-7}$& $4.53\times 10^{-7}$\\
\hline
 $D_{s}^{+}\rightarrow K^{+} V$ & 
 $1.74\times 10^{-7}$&  
 $1.04\times 10^{-7}$& 
 $8.26\times 10^{-8}$ & 
 $4.47\times 10^{-6}$& 
 $5.99\times 10^{-7}$& - \\
\hline
\end{tabular}
\end{center}
\caption{ $G_{D_s\pi V}$ and $G_{D_s K V}$ dimensionless couplings.}
\label{couplings1}
\end{table}

The $g_V$ coupling in Eq.~(\ref{mvector}) can be obtained from the $V\to l^+l^-$ decay, considering the amplitude in the vector dominance approach, by
\begin{equation}
\mathcal{M}= \frac{e^2}{g_V}  \bar u_l \gamma^\mu v_l \epsilon_\mu ,
\end{equation}
where $\epsilon_\mu$ is the vector polarization tensor and $ \bar u_l$ and  $v_l$ are the  spinors for the corresponding leptons. Thus,
\begin{equation}
g_{V}=\Big [
\frac{4\pi \alpha^2 (2m^2_l+m^2_V)(m^2_V-4m^2_l)^{1/2}}
{3\Gamma_{Vll}m^2_V} 
\Big ]^{1/2},
\end{equation}
where $\Gamma_{Vll}$ is the decay width of the $V$ meson into a pair of leptons.
In Table~\ref{tablagv} we show the values obtained from an analysis considering the muon and electron decay modes weighted averages~\cite{ourcouplings}. Given the lack of experimental information for the $\rho (1450)$ meson, we only quote a central value as reference but its contribution is not considered in this analysis.  

\begin{table}[t]
\begin{center}
\begin{tabular}{|c|c|c|c|c|}
\hline
$V\to$ & $\rho$ & $\omega$ &$\phi$  & $\rho (1450)$ \\
\hline
$g_V$ & $4.97 \pm 0.02$ & $16.97\pm0.3$ & $13.53 \pm 0.34$  & $13.53$\\
\hline
\end{tabular}
\end{center}
\caption{ $g_{V}$ dimensionless couplings obtained from weighted averages of electron and muon decay modes.}
\label{tablagv}
\end{table}
 
With the aforementioned information at hand, we consider the contribution from the known $\rho$, $\omega$ and $\phi$ vector resonances to the $D^{+}_{s}\rightarrow \pi^{+}\ell^{-}\ell^{+}$ decay as in Eq.~(\ref{mvector}).
The amplitude can be set as
\begin{eqnarray}
\mathcal{M}_{LD}&=& i e^2(P+p_3)_\nu l^\nu \sum_{V=\rho,\omega,\phi}\frac{G_{D_s\pi V}}{ g_V}\frac{1}{k^2-m_V^2+im_V\Gamma_V}\nonumber\\
&=&  i e^2 a_\phi e^{i \delta_\phi}
\bigg (
\frac{a_{\rho\phi} e^{i \delta_{\rho\phi}}}{k^2-m_\rho^2+im_\rho\Gamma_\rho}
+  \frac{a_{\omega\phi}e^{i \delta_{\omega\phi}}}{k^2-m_\omega^2+im_\omega\Gamma_\omega}
\nonumber\\
&&+\frac{1}{k^2-m_\phi^2+im_\phi\Gamma_\phi}
\bigg ) (P+p_3)_\nu l^\nu,
\label{mld}
\end{eqnarray}
where we factorize the $\phi$ parameters and make explicit the phases, such that $e^{i\delta_{V\phi}}$ is the relative phase between the vector mesons $V$ and $\phi$,  $a_\phi e^{i \delta_\phi} \equiv \frac{G_{D_s\pi \phi}}{g_{\phi}}$ and $a_{V\phi}  \equiv |\frac{G_{D_s\pi V}}{g_{V}}/ \frac{G_{D_s\pi \phi}}{g_{\phi}}|$, where $V=\rho \ , \ \omega$. Using the values in Tables \ref{couplings1} and \ref{tablagv}, we obtain $a_\phi=7.63\times 10^{-8}$ and $a_{\rho\phi}=0.11$. The latter is comparable with $0.13 $ obtained by considering a different parametrization as in Ref.~\cite{Bause:2019vpr}. The $SU(3)$ flavor symmetry establishes a relationship between $g_{\omega}$ and $g_{\rho}$ to be $g_{\omega}=3g_{\rho}$, which  can be compared with the experimental values (see Table~\ref{tablagv}) obtained from their decays into two leptons, $g_{\omega}=3.41 g_{\rho}$~\cite{ParticleDataGroup:2020ssz}. Based on this observation, we assume that this symmetry is a good approximation and identify the relative phase between the $\rho$ and $\omega$ from the $\bar{d}d \sim -\frac{1}{\sqrt{2}}\rho^0+\frac{1}{\sqrt{2}}\omega$ current, associated to the production of such mesons \cite{Fajfer:2005ke}, namely $e^{i\delta_{\omega\phi}} = - e^{i\delta_{\rho\phi}}$. This assumption may have implications that should be investigated. We incorporate $SU(3)$ symmetry breaking through the experimentally extracted couplings. In summary, the only unknown parameter is the relative phase $\delta_{\rho\phi}$ and it drives most of the error estimates of the hadronic contribution~\cite{Gisbert:2020vjx,deBoer:2015boa,Bause:2019vpr,Fajfer:2015mia,Burdman:2001tf,Fajfer:2005ke}. 

\begin{figure}
 \centering
\includegraphics[width=10cm,angle=-90]{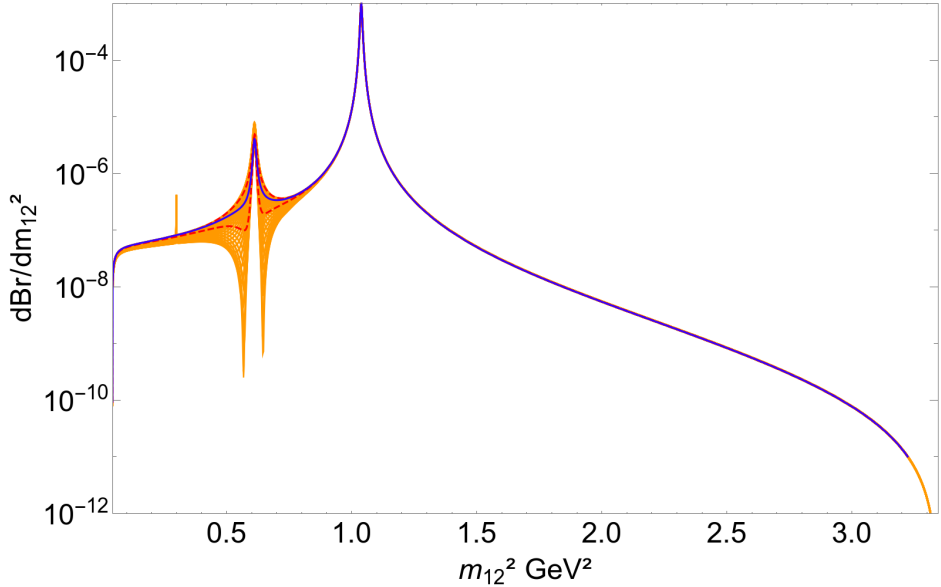}
\caption{Di-muon invariant mass ($m_{12}^2$) for the $D_s \to \pi \mu\mu$ decay obtained from LD contributions.
The wide shaded area corresponds to the region allowed by the relative phase $\delta_{\rho \phi}$ when it is varied in the domain, while  the dashed lines delimit the $1\sigma$ region obtained from the fit of $\delta_{\rho \phi}$ to LHCb data~\cite{LHCb:2013hxr}. The solid line corresponds to the central fit value.
}
\label{Fig:2}
\end{figure}

For the $D^{+}_{s}\rightarrow \pi^{+}\mu^{-}\mu^{+}$ decay, the di-lepton invariant mass has been measured by the LHCb experiment~\cite{LHCb:2013hxr}, which can be used to determine $\delta_{\rho\phi}$. The experiment does not provide the details on the resonances analysis, which should include the corresponding phases. Thus, in order to illustrate the feasibility to extract $\delta_{\rho\phi}$, we have computed the di-lepton invariant mass using Eq. (\ref{mld})  and performed a $\chi^2$ fit in the region between the $\rho$ and the $\phi$ resonances. In this particular case, the $a_\phi$ and $a_{V\phi}$ are effective factors that take into account  all  detection effects, including efficiencies which are not provided by the experiment, and their magnitudes are fixed to the individual contributions at the nominal mass of the corresponding mesons in the data, leaving the relative phase as the only free parameter. The favored relative phase is $\delta_{\rho\phi}=(0.44 \pm 0.21) \pi$, where the uncertainty accounts for the fitting uncertainty, in addition to a conservative 10\% due to the data extraction procedure from Figure 4 in Ref.~\cite{LHCb:2013hxr}. The obtained value should be considered as an indication of the experimental feasibility to determine this parameter accurately with a proper handling of the data, with a thorough incorporation of the detector capabilities. 
In Fig.~\ref{Fig:2}, we exhibit the role of the phase in the di-muon spectrum for the $D_s \to \pi \mu\mu$ decay normalized to the $D_s$ total width (denoted by $dBr/dm_{12}^2\equiv \frac{1}{\Gamma_{D_s}}d\Gamma/dm_{\mu\mu}^2$), for both the full range and the one from the fit. 

Pseudo-scalar mesons like the $\eta$ and $\eta^\prime$ also contribute in the low invariant mass regime. 
The process can be identified, using Eq. (\ref{weakl}) effective interaction, with the pseudo-scalar $D_s$ to $\pi$ transition that then produces the $\eta$ meson. The last emits the lepton pair in $s$-wave 
($\bar{l}\gamma^5 l$ effective current).
In Fig.~\ref{Fig:1}b we show the $\eta$ mediated decay mechanism \cite{Gan:2020aco,Landsberg:1985gaz, Bergstrom:1983ay, Ametller:1993we,Bergstrom:1982zq,Ametller:1997uk,Pratap:1972tb,Gan:2020aco}.
The amplitude is usually described considering a Breit-Wigner parametrization, with a single effective coupling ($a_\eta$) determined from experimental information, of the form 
 \begin{equation}
C_P=i  
\frac{a_\eta}{k^2-m_\eta^2+i\epsilon}.
\end{equation}
Here, we consider the $D_s \to \pi\eta$ \cite{Rosner:1999xd,Huong:2016gob} and $\eta \to \mu \mu$ decay widths to estimate the corresponding couplings. Then,
$a_\eta$ is related to the partial decay effective couplings by $a_{\eta}=g_{D_s \pi\eta}g_{\eta\mu\mu}$, where $g_{D_s \pi\eta}=1.54\times 10^{-6}$~GeV and $g_{\eta\mu\mu}=1.94\times 10^{-5}$. A similar description corresponds to the $\eta^\prime$ contribution. The narrow widths of both the $\eta$ and $\eta^\prime$ resonances bring them to play a negligible role in the di-lepton spectrum, and we keep only the $\eta$ contribution to illustrate this. In Fig.~\ref{Fig:2} the narrow distribution of the $\eta$ resonance is slightly visible at the current scale.
 
The non-resonant region, for the lepton pair invariant mass distribution, is defined by the LHCb experiment as the one below the $\eta$ and above the $\phi$ resonances~\cite{LHCb:2013hxr}. The bounds for the branching ratios in the search for new physics, in processes involving FCNC contributions, are estimated using the corresponding experimental data and considering $D_s^+ \rightarrow \pi^+ \phi(\phi \to \mu^+\mu^-)$ as the normalization channel. 
Next, we explore the role of the $\delta_{\rho\phi}$ uncertainties as we move away from the $\phi$ resonance. For that purpose, we define three kinematical regions to estimate the corresponding contribution to the branching ratio: the central region, defined by four times the  $\phi$ decay width ($\Gamma_{\phi}$) around the $\phi$ mass,  $[m_{\phi}-4\Gamma_{\phi} \ , \ m_{\phi}+4\Gamma_{\phi}] $;
 and the other two complementary regions to cover the full phase space below and above the central region,
 $[2m_{\mu} \ ,\ m_{\phi}-4\Gamma_{\phi}] $ and $[m_{\phi}+4\Gamma_{\phi},m_{D_{s}}-m_{\pi}]$. 
Table \ref{regions} shows the results for the branching ratios in each region, where in the first row we report the range for the LD part, corresponding to the variation of the $\delta_{\rho\phi}$ phase in its total allowed range, and in the second row the branching ratios with the corresponding uncertainties for the $\delta_{\rho\phi}$ phase as obtained from  the fit to the LHCb data.
We also include the corresponding branching ratios for the SD contributions, which will be discussed later, and the LD-SD interference for the $\delta_{ \rho\phi}$ central value.

\begin{table}[htbp]
\begin{center}
\begin{tabular}{|c|c|c|c|c|}
\hline
\hline
 {$\mathcal{B}$ } &{\small $[m_{\phi}-4\Gamma_{\phi},m_{\phi}+4\Gamma_{\phi}]$ }& {\small$[2m_{\mu},m_{\phi}-4\Gamma_{\phi}]$ }  &  {\small $[m_{\phi}+4\Gamma_{\phi},m_{D_{s}}-m_{\pi}]$} & {\small Total} \\
\hline
LD$_{full \ \delta_{\rho\phi}}$ & $[1.2371,1.2384]\times 10^{-5}$ &  $[7.97,8.08]\times 10^{-7}$ & $[4.46, 4.52]\times 10^{-7}$& $[1.36, 0.003]\times 10^{-5}$\\
\hline
LD$_{fit \ \delta_{\rho\phi}}$ & $(1.2378 \pm0.0004)\times 10^{-5}$ &  $(8.05\pm0.04)\times 10^{-7}$ & $(4.46\pm 0.02)\times 10^{-7}$& $(1.362\pm 0.001)\times 10^{-5}$\\
\hline
{ SD} & $8.726\times 10^{-17}$&  $2.0\times 10^{-15}$& $8.084\times 10^{-16}$ &$2.896\times 10^{-15}$ \\
\hline
{LD-SD} & $5.706\times 10^{-13}$&  $6.31\times 10^{-11}$& $2.4\times 10^{-11}$& $3.961\times 10^{-11}$  \\
\hline
\hline
\end{tabular}
\end{center}
\caption{Branching ratio ($\mathcal{B}$) contributions in the three mass regions as defined in the text.}
\label{regions}
\end{table}

\section{Angular observables}
\begin{figure}
 \centering
\includegraphics[width=12cm,angle=-90]{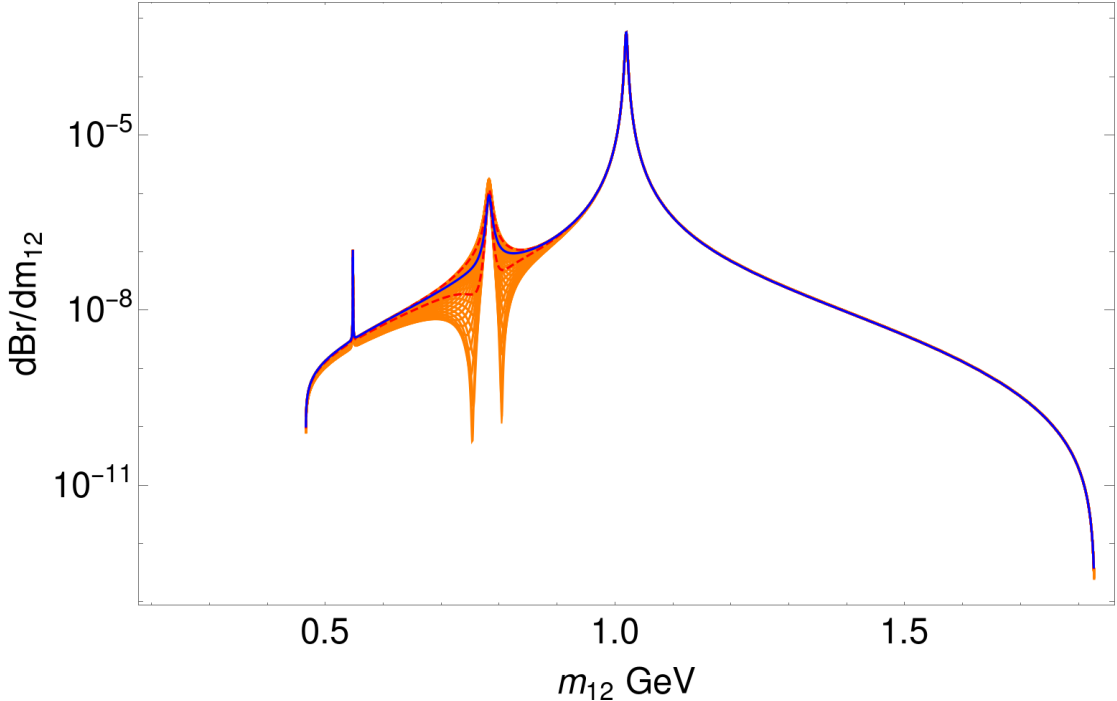}
\caption{Di-muon invariant mass distribution at $\cos \theta=0.1$ (denoted by $dBr/dm_{12} \equiv \frac{1}{\Gamma_{D_s}}d\Gamma/dm_{\mu\mu}$)  obtained from LD contributions, as an example in the most kinematically favored region. The shaded area corresponds to the region allowed by the relative phase $\delta_{\rho\phi}$ when it is varied in its full domain,
while the dashed lines delimit the 1$\sigma$ region obtained from the fit of $\delta_{\rho\phi}$ to LHCb data. The solid line corresponds to the central fit value.
}
\label{Fig:a1}
\end{figure}

The lack of information on the relative phase $\delta_{\rho\phi}$ is the main source of the uncertainty on the LD description of the $D^{+}_{s}\rightarrow \pi^{+}\ell^{-}\ell^{+}$ decay. An observable insensitive to it would be useful in order to consider the full kinematical region, i.e. without excluding the resonant region. For that purpose, we have explored  the invariant mass ($M_{inv}$) distribution of the lepton pair at a given angle of one of the leptons emission with respect to the pion  ($\cos\theta$). This can be written as follows:

\begin{equation}
\frac{d\Gamma}{dM_{inv} \ d\cos\theta }=\int \frac{|\mathcal{M}|^2 \delta(M_{D_s}-E_1-E_{l_1}-E_{l_2})}{8M_{D_{s}}^{2} (2\pi)^3} \frac{|\textbf{p}_1||\textbf{l}_1| |M_{inv}|}{E_{l_2}} dE_{l_1},\label{dGdMdcos}
\end{equation}
where 
\begin{equation}
E_{l_2}=\sqrt{m^2_l+ \textbf{p}_1^2+\textbf{l}_1^2+ 2 |\textbf{p}_1 || \textbf{l}_1 | \cos\theta}, \
E_1=(M^2_{D_s}+m_\pi^2-M_{inv}^2)/2M_{D_s}.
\end{equation}
The integration on $dE_{l_1}$, using the $\delta$ function, gives the angular distribution of the invariant mass. In Fig.~\ref{Fig:a1}, we plot the lepton pair invariant mass distribution at $\cos\theta=0.1$ (Eq.~\ref{dGdMdcos}), as an example in the most kinematically favored region. The relative phase within its full domain (shaded area) is compared with the restricted region from the fit, showing the advantages of the determination of the phase already at this stage. The same is true for any other angular value, although the invariant mass is modified by the kinematical restrictions.

Angular observables have been considered to identify asymmetries related to CP-violating terms, by integrating out the angular dependence \cite{Mahmoudi:2012uk,deBoer:2015boa,Gisbert:2020vjx,Bause:2019vpr,Fajfer:2015mia,Sahoo:2017lzi,Bharucha:2020eup,Burdman:2001tf,Paul:2011ar}.
Here, we limit the forward-backward distribution to a specific angle, $\cos\theta$ as defined above, by:
\begin{equation}
A_{FB}|_{\cos\theta}=
\frac
{\frac{d\Gamma}{dM_{inv}d\cos\theta}|_{\cos\theta}  -
\frac{d\Gamma}{dM_{inv}d\cos\theta}|_{-\cos\theta}
}
{\frac{d\Gamma}{dM_{inv}d\cos\theta}|_{\cos\theta}  +
\frac{d\Gamma}{dM_{inv}d\cos\theta}|_{-\cos\theta}
}.
\end{equation}

\begin{figure}
 \centering
\includegraphics[width=7cm,angle=-90]{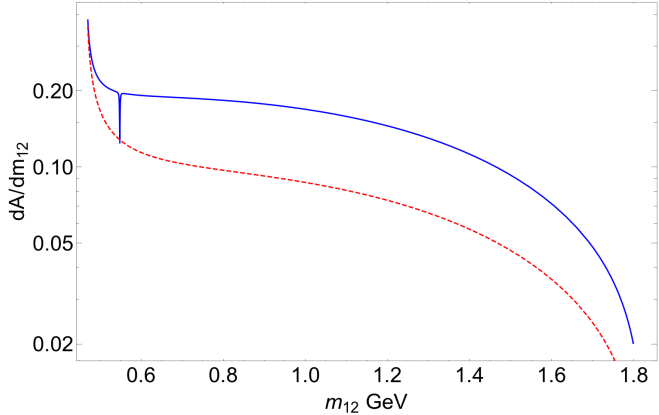}
\caption{$dA/m_{12}=A_{FB}|_{cos\theta}$ distribution for $\cos \theta=0.1$ (solid line) obtained from LD contributions, compared to the pure phase space estimation (dashed line).}
\label{Fig:a2}
\end{figure}

In Fig.~\ref{Fig:a2}, we show the $A_{FB}|_{\cos\theta}$ (denoted by $dA/dm_{12}$) distribution for  $\cos \theta=0.1$ (solid line). We observe that the vector resonance region loses its typical features, except around the $\eta$ mass. Even though the relative phase is varied within its full domain, we observe no visible dependence on it. The interference term between the $\rho$ and the $\phi$ mesons, which carries the phase dependence, is not null. The combination of the relative magnitudes of the resonances drives this interference to become relatively small compared to the full contribution.
The shape is similar to that of a non-resonant background component, and could be used as a reference for other contributions. This distribution can be compared with the corresponding result when considering a pure phase space (PS) approach  (dashed line)  which follows a similar behavior, with a  relative area under the curve of $A_{FB}^{(PS)} |_{\ \cos\theta} /A_{FB} |_{\ \cos\theta}  = 0.1024 / 0.1863 =  0.55$ for $\cos\theta=0.1$.

\subsection{Background}
\begin{figure}
  \centering
\includegraphics[width=9cm,angle=-90]{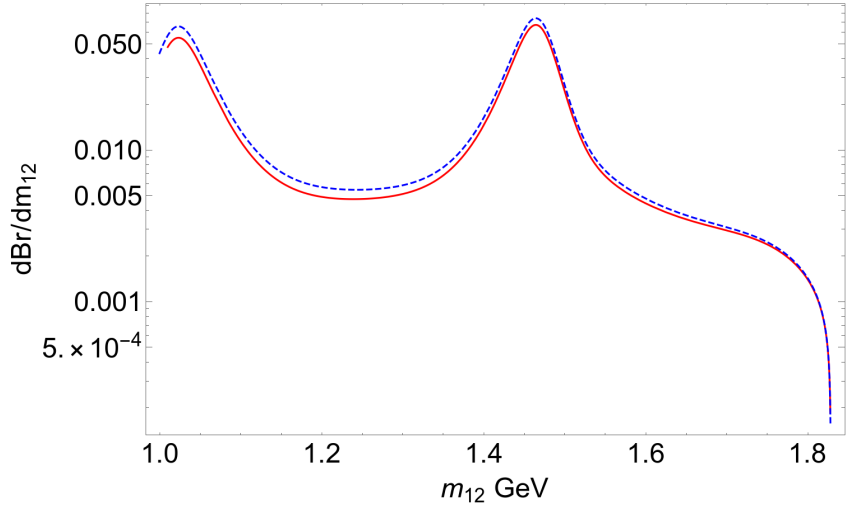}
\caption{ Di-pion invariant mass distributions at given angles, $\cos \theta=0.1$ (solid line) and $ -0.1$  (dashed line), for the $D_s \to 3\pi$ decay. }
\label{Fig:AngularDs3pi}
\end{figure}

\begin{figure}
  \centering
\includegraphics[width=7cm,angle=-90]{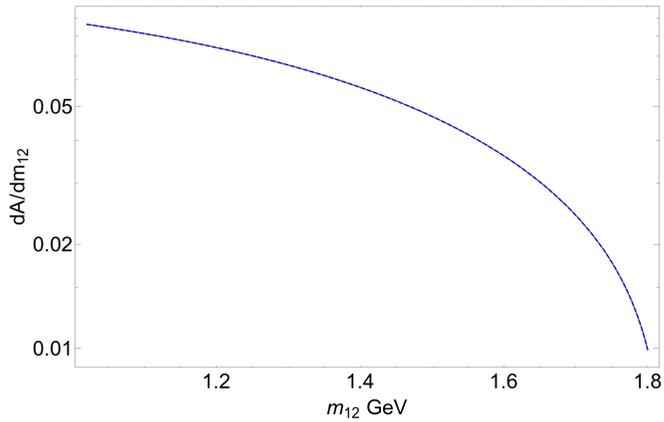}
\caption{$A_{FB}|_{\cos\theta}$ distribution at $\cos \theta=0.1$ for the $D_s \to 3\pi$ decay. 
No difference is exhibited compared to the pure phase space estimation.}
\label{Fig:FBDs3pi}
\end{figure}

The main background for the $D^{+}_{s}\rightarrow \pi^{+}\ell^{-}\ell^{+}$ decay is given by the $D_s \to \pi \pi \pi$ decay which then can produce either the $\mu^+\mu^-$ or $\mu^+\mu^+$ mode through the decay of the corresponding pions. This process involves intermediate  mesons producing a resonant invariant mass. In order to explore its behavior in the angular observables, we model this background as a sum of Breit-Wigner ($BW$) resonances using the experimental information for the relative magnitudes ($a_i$) and phases ($\delta_i$) from the FOCUS collaboration~\cite{FOCUS:2003tdy,ParticleDataGroup:2020ssz}, whose analysis includes the $f_0(980)$, $f_0(1300)$, $f_0(1200-1600)$, $f_0(1500)$ and $f_0(1750)$ states. These components have also been analysed  recently by the BESIII collaboration~\cite{BESIII:2021jnf}. Studies on the nature of such resonances \cite{Sakai:2017iqs,Dias:2016gou,Sekihara:2015iha} are relevant, but out of the scope of our work. Then, the amplitude can be set as
\begin{equation}
\mathcal{M}_B= a_0e^{i\delta_0} +\sum_{i} a_i e^{i\delta_i} BW_i,
\end{equation}
where $a_0e^{i\delta_0}$ is the non-resonant contribution, and the sum is the contribution of the $s$-wave states mentioned above.
In Fig.~\ref{Fig:AngularDs3pi}, we show the di-pion invariant mass distribution at given angles, $\cos \theta=0.1$ and $ -0.1$ (solid and dashed lines respectively)  in a similar way as in the semi-leptonic case. 
Also, in Fig.~\ref{Fig:FBDs3pi}, we show the $A_{FB}|_{\cos\theta}$ distribution at $\cos \theta=0.1$ for the $D_s \to 3\pi$ decay. No difference is exhibited compared to the pure phase space.

\subsection{Short distance contribution}
\begin{figure}
 \centering
\includegraphics[width=15cm]{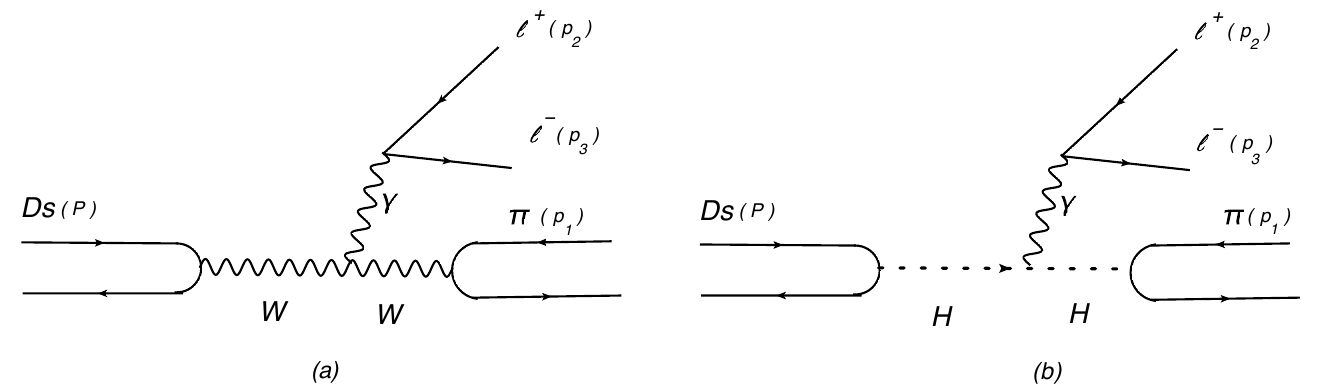}
\caption{Short distance (SD) contributions to the $D_s \to \pi \mu\mu$ decay from the (a) SM and (b)  2HDM-II.}
\label{Fig:SD}
\end{figure}

The SD contribution comes from the decay of the $D_s$ into a $W$, which emits a photon that produce a pair of leptons, and then the final $W$ becomes a pion, as depicted in Fig.~\ref{Fig:SD}(a).  In the SM, the dimension six effective Lagrangian gives rise to the $WW\gamma$ interaction \cite{BUCHMULLER1986621,Grzadkowski:2010es}
\begin{equation}
    \mathcal{L}_{WW\gamma}=ie \Gamma_{\alpha\rho\theta}A^\alpha W^\rho W^\theta,
\end{equation}
where $A^\alpha$, $W^\rho$ and $W^\theta$ are the photon and $W$ fields respectively.
$\Gamma_{\alpha\rho\theta}=g_{\alpha\rho}(-k-P)_{\theta}+g_{\theta\alpha}(k-p_{1})_{\rho}+g_{\rho\theta}(P+p_{1})_{\alpha}$ is the $WW\gamma$ vertex \cite{HAGIWARA1987253}. 
The weak transition matrix element is
\begin{equation}
   \bra{\gamma^*(k,\alpha)\pi(p_1) } \mathcal{H}_{eff}\ket{D_s(P)}
   =\bra{\pi }\mathcal{H}_{weak}\ket{W} 
    \bra{\gamma^*(k,\alpha)W}\mathcal{L}_{WW\gamma}\ket{W}
    \bra{W}\mathcal{H}_{weak}\ket{D_s}.
\end{equation}
 By coupling the leptonic current to the photon the amplitude is set as
\begin{eqnarray}
\mathcal{M}_{SD}=-i\frac{G_{F}V^{\ast}_{cs}V_{ud}f_{D_{s}}f_{\pi}e^{2}}{\sqrt{2}M^{2}_{W}k^{2}} P^{\rho}p^{\theta}_{1}
l^{\alpha} \Gamma_{\alpha\rho\theta},
\end{eqnarray}
where the weak transition matrix element for the pseudo-scalar meson $P$ is $\bra{ 0} \bar{u}\gamma_{\mu}\gamma_{5}d
 \ket{P}  = i f_P p_{P}^\mu$, $u$ and $d$ denoting up and down-like quarks, and $f_\pi=0.13$ GeV and $f_{D_s}=0.249$ GeV~\cite{ParticleDataGroup:2020ssz,Rosner:1999xd}.

\begin{figure}
 \centering
\includegraphics[width=7cm,angle=-90]{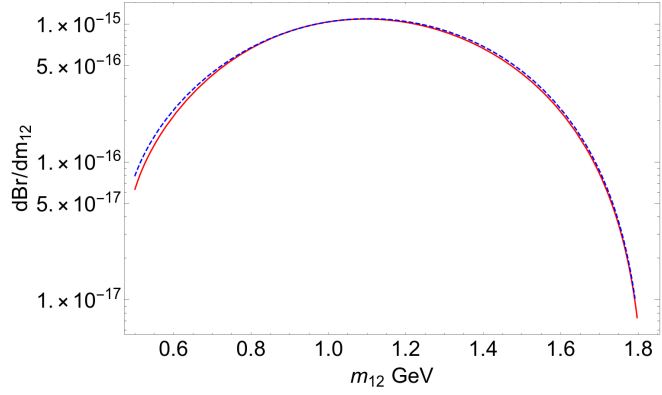}
\caption{Di-muon invariant mass at given angles, $\cos \theta=0.1 $ (solid line) and $ -0.1$ (dashed line) from the SD transition mechanism.}
\label{SD:AngularDspi}
\end{figure}

\begin{figure}
 \centering
\includegraphics[width=7cm,angle=-90]{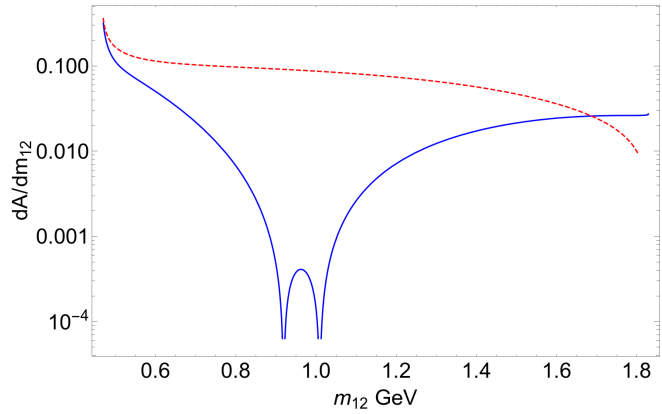}
\caption{SD $A_{FB}|_{\cos\theta}$ distribution for $\cos \theta=0.1$ (solid line), compared to the pure phase space estimation (dashed line).}
\label{SD:FBDspi}
\end{figure}

\noindent In Fig.~\ref{SD:AngularDspi}, we show the di-muon invariant mass at given angles $\cos \theta=0.1 $ (solid line) and $ -0.1$  (dashed line), from the SD transition mechanism. In Fig.~\ref{SD:FBDspi},  we show the $A_{FB}|_{\cos\theta}$ distribution for $\cos \theta=0.1$ (solid line), compared to the pure phase space (dashed line). This case exhibits a different behavior compared not only to the phase space, but also to the LD result, as seen in Fig.~\ref{Fig:a2}.
 
\subsection{Non-standard contributions}  
\begin{figure}
 \centering
\includegraphics[width=8cm,angle=-90]{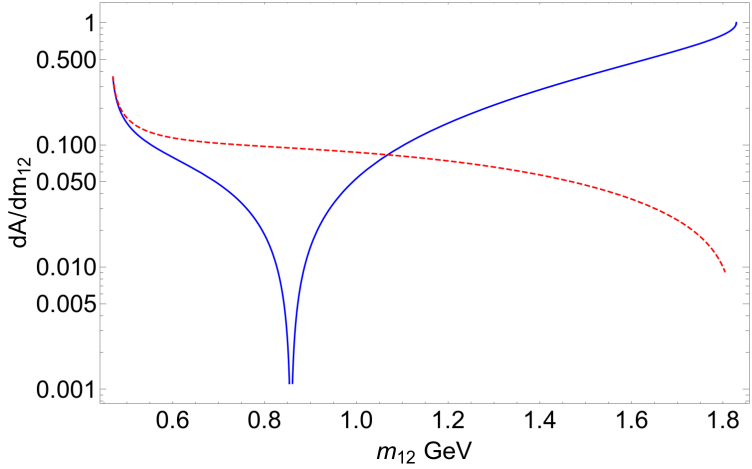}
\caption{$A_{FB}|_{\cos\theta}$ distribution for the 2HDM-II at $\cos \theta=0.1$ (solid line), compared to the pure phase space estimation (dashed line). }
\label{Fig:2hdm}
\end{figure}

Although the $D^{+}_{s}\rightarrow \pi^{-}\ell^{+}\ell^{+}$ decays is not an ideal scenario to look for non-standard contributions, we can consider the process driven by a charged Higgs, as it would come from the 2HDM-II model \cite{Hou:1992sy,Branco:2011iw}, to show the behavior of the observable considered in this work. Following a similar approach as in the SD case, the effective Lagrangian interaction for the Higgs with the photon is taken as in $sQED$
\begin{equation}
    \mathcal{L}_{HH\gamma}=ie ((\partial_\mu \phi^*)\phi-(\partial_\mu \phi)\phi^*)A^\mu,
\end{equation}
where $\phi$ is the Higgs field.
The transition matrix element is set as
\begin{equation}
   \bra{\gamma^*(k,\mu)\pi(p_1) } \mathcal{H}^{eff}_{2HDM}\ket{D_s(P)}
   =\bra{\pi }\mathcal{H}_{2HDM}\ket{\phi} 
    \bra{\gamma^*(k,\mu)\phi}\mathcal{L}_{HH\gamma}\ket{\phi}
    \bra{\phi}\mathcal{H}_{2HDM}\ket{D_s}.
\end{equation}
By coupling the leptonic current to the photon, the amplitude is then
\begin{equation}
\mathcal{M}_H =
\frac{i e^2G_F}{\sqrt{2}}V_{ud}V^{\ast}_{cs}f_\pi f_{D_s}
m_\pi^2 m_{D_s}^2
\frac{1}{\tan^2\beta} \frac{1}{k^2m_H^4}P_\mu l^\mu,
\end{equation}
where, in order to compute the hadron-Higgs transition, the general coupling of the quarks to the  charged Higgs is of the form
\begin{equation}
    \frac{-\sqrt{2}V_{ud}}{v}(m_u \cot\beta P_L+ m_d \tan\beta P_R)dH^+,
\end{equation}
for $u$ and $d$-like quarks. The transition matrix element of the pseudo-scalar coupled to the charged Higgs is taken as 
$
\bra{0}\bar{u}\gamma_{5}d \ket{P^{-}}=-if_{P}(m^{2}_{P}/m_{u})
$ \cite{Hou:1992sy}.
We ignore $m_{s}$ compared to $m_{c}$ for the $D_s$ and for simplicity we keep only the $u$ contribution for the pion. 
For illustration, the parameters from the model are taken as $m_H=600$ GeV and $\tan\beta=10$, which are within the current experimental bounds \cite{Arbey:2017gmh,Haller:2018nnx,ParticleDataGroup:2020ssz}. In Fig.~\ref{Fig:2hdm}, we plot the $A_{FB}|_{\cos\theta}$ distribution for $\cos \theta=0.1$ (solid line). Note that for this observable the particular values of $m_H$ and $\tan\beta$ cancel out. We observe that the distribution is quite different from the pure phase space (dashed line) and, therefore, exhibits a clear signature. The dip position depends on the relative mass of the pion with respect to the $D_s$.

\section{Discussion}
The new physics scenarios that may be at reach in the charm sector require a proper account of the different hadronic decay modes. For instance, a better knowledge of the full aspects of non-FCNC processes would help to identify the truly FCNC features. In particular, the $D^{+}_{s}\rightarrow \pi^{+}\ell^{-}\ell^{+}$ decay has been used as a reference channel in the understanding of suppressed modes involving FCNC. The process is dominated by hadronic resonances that decay to $\ell^+\ell^-$, from the $\eta$ to the $\phi$ meson. This region is usually accounted in a phase space approach due to the large uncertainties associated, which has implications on the data analysis, namely efficiencies may be largely affected. We explored the role of the dominant source of uncertainty, that is, the relative strong phase between the $\rho$ and the $\phi$ mesons. An analysis of the LHCb data for the di-muon invariant mass in the $D^{+}_{s}\rightarrow \pi^{+}\mu^{-}\mu^{+}$ decay provides a first approach to this phase, namely $\delta_{\rho\phi}=(0.44 \pm 0.24) \pi$. Current LHCb dataset can be used to obtain a more accurate value of the $\delta_{\rho\phi}$ phase, by an improved analysis of the di-muon invariant mass in a wider region, taking into account all detector effects missed in this work. 

Observables insensitive to the $\delta_{\rho\phi}$ phase would be also useful to keep the hadronic contributions under control. We studied the invariant mass of the lepton pair at a given angle of one of the leptons emission with respect to the pion. 
We then computed the corresponding forward-backward distribution and found that it exhibits no dependence on this parameter. The resonance region is smeared, although global effects are observed by comparing with the pure phase space approach. 
In order to consider this observable in more general scenarios, that would allow the experimentalists to disentangle standard (LD and SD) and non-standard contributions, we  analysed the behavior for the resonant background process $D^{+}_{s}\rightarrow \pi\pi\pi$, the SD contribution in the SM, and the charged Higgs mediated process as given by the 2HDM-II.  We found that they exhibit distinguishable features among themselves, which might be useful in the understanding of the different contributions. Similar behaviors are expected for the $D_s \to \pi e^+e^-$ case. 

The Belle II experiment will collect an integrated luminosity of about 50~ab$^{-1}$ in the next few years, and has an ambitious charm sector program~\cite{Belle-II:2018jsg}. With such a large dataset, we also expect that Belle II may be capable to provide precise measurements of the proposed observable, with reduced systematic uncertainties given the low background environment in $e^+e^-$ collisions, almost unbiased selections, and excellent particle identification.

We would like to conclude by pointing out  that the usually excluded resonant region in the search for non-SM signals may be considered, provided such region is under theoretical control, and in this work we explored possible observables that can help to reach that goal.

\section*{Acknowledgements}
We acknowledge the support of DGAPA-PAPIIT UNAM, grant no. IN110622, M. S. acknowledges financial support from CONACyT through grant 332429.

\printbibliography

\end{document}